\documentclass[%
 reprint,
superscriptaddress,
nofootinbib,
 amsmath,assume,
 aps,
 prl,
floatfix,
]{revtex4-2}

\usepackage{graphicx}
\usepackage{dcolumn}
\usepackage{bm}

\usepackage{enumitem}
\usepackage{epsfig}
\usepackage{bm}
\usepackage{amsmath}
\input{epsf}
\usepackage{psfrag}
\usepackage{subcaption}
\usepackage{xcolor}
\usepackage{pdfpages}
\usepackage[normalem]{ulem}
\makeatletter
\AtBeginDocument{\let\LS@rot\@undefined}
\makeatother
\newcommand{\bea}{\begin{eqnarray}}
\newcommand{\eea}{\end{eqnarray}}

\newcommand{\nn}{\nonumber}

\newif\ifshowcomments
\showcommentstrue 

\begin{document}

\title{TMDs from Semi-inclusive Energy Correlators}

\author{Xiaohui Liu}
 \email{xiliu@bnu.edu.cn}
 \affiliation{Department of Physics, Beijing Normal University, Beijing, 100875, China}
 \affiliation{Key Laboratory of Multi-scale Spin Physics, Ministry of Education, Beijing Normal University, Beijing 100875, China}

\author{Hua Xing Zhu}%
 \email{zhuhx@pku.edu.cn}
\affiliation{School of Physics, Peking University, Beijing, 100871, China}%
\affiliation{Center for High Energy Physics, Peking University, Beijing 100871, China}%

\begin{abstract}
We introduce a novel category of observables known as the Semi-Inclusive Energy Correlators (SIECs), an extension of  
the recently proposed nucleon energy correlator to integrate a new element, the fragmenting energy correlation function. These SIECs gauge the correlation between the examined hadron and the surrounding radiations, providing a comprehensive tomography of the radiative patterns originating from initial state radiation or parton fragmentation. As such, they could function as the generating functions for numerous kinematic distributions. To illustrate, we 
find a direct relation between the transverse momentum moments (TMMs) of the transverse momentum-dependent (TMD)  distributions and the SIECs. We demonstrate how the TMMs of the TMD parton distributions and the fragmentation functions can be distinctively derived from the nucleon energy correlator and the fragmenting energy correlator, respectively, without enforcing the back-to-back kinematics. 
\end{abstract}

\maketitle
 
\textbf{\emph{{\color{magenta}Introduction.}}}
%
%
The femtoscale structure of the nucleon has long been a primary focus of nuclear physics. The upcoming QCD facilities~\cite{AbdulKhalek:2021gbh, Proceedings:2020eah,Anderle:2021wcy} are expected to significantly enhance our understanding of the partonic structure of the nucleon and nucleus. The investigation of the nucleon and nucleus tomography has involved examining their transverse momentum-dependent (TMD) structure functions~\cite{Boussarie:2023izj}, 
with a focus on the current fragmentation region, 
which has yielded remarkable progress in nucleon tomography. Meanwhile, extension into the target fragmentation and central~(soft) region has also been emphasized recently (see~\cite {Boglione:2016bph,Chen:2024brp} and references therein for recent developments), pioneered by Trentadue and Venezlano, who introduced the fracture functions~\cite{Trentadue:1993ka} that measured the longitudinal momentum distribution of probed hadron in the target fragmentation region. 


Recently, the nucleon Energy-Energy Correlator~(EEC) has been proposed by the authors~\cite{Liu:2022wop} to probe nucleon tomography through measurement in the target fragmentation region. The nucleon EEC suggests probing the nucleon structure by directly measuring the energy-weighted angular correlation inside the beam remnant jet. This innovative approach has found practical applications in probing gluon saturation and the polarized gluons within nucleons~\cite{Liu:2023aqb,Li:2023gkh}. 

\begin{figure}
  \begin{subfigure}{.5\linewidth}
    \centering
    \includegraphics[width=\linewidth]{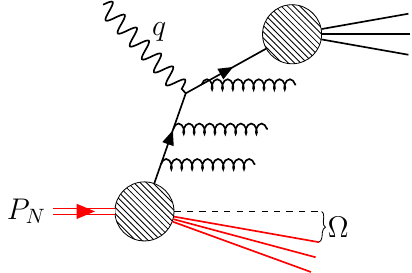}
    \caption{
    Nucleon energy correlator. The measurement records the energy flow at angle $\Omega$ in the 
    target fragmentation region.}
    \label{fg:sub1}
  \end{subfigure}%
    \begin{subfigure}{.5\linewidth}
    \includegraphics[width=\linewidth]{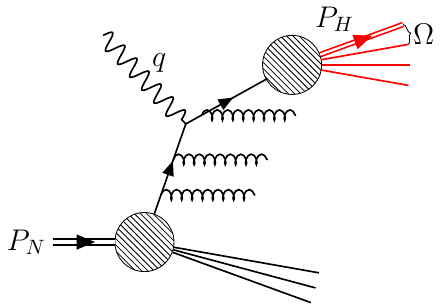}
    \caption{
    Fragmenting energy correlator. One records energy flow at angle $\Omega$ surrounding the tagged $h$ in the
    jet fragmentation region.}
    \label{fg:sub2}
  \end{subfigure}
  \caption{SIEC with a single energy weighting measured in the target (a) and the jet (b) fragmentation region. 
  }
  \label{fg:array}
\end{figure}

In this work, we introduce the concept of Semi-Inclusive Energy Correlators~(SIECs), which encompasses the nucleon EEC as a special case, meanwhile extends into the current or jet fragmentation region. Specifically, the SIEC measures energy-weighted angular correlation among an examined 
hadron 
and the radiations surrounding it. The correlation can either be measured in the target fragmentation region, where the examined hadron represents the struck target nucleon, Fig.~\ref{fg:sub1}; or be measured in the jet or current fragmentation region, in which case the studied hadron denotes a final-state identified hadron $h$, Fig.~\ref{fg:sub2}. 


The angular correlation of the SIECs is inherently associated with the rotational structure of the underlying non-perturbative mechanism, presenting a fresh approach to studying the non-perturbative spin structure of hadrons, such as the Sivers and Collins effects~\cite{Sivers:1989cc,Collins:1992kk}. Additionally, the comprehensive information regarding the radiations contained within SIECs allows for the direct construction of numerous kinematic distributions, such as the transverse momentum moments (TMMs)~\cite{delRio:2024vvq} of the TMD distributions, through the measurement of SIECs, and hence provides a distinctive probe to these crucial distributions, as we will demonstrate in the rest of the work.

\begin{figure}
    \centering
    \includegraphics[width=0.7\linewidth]{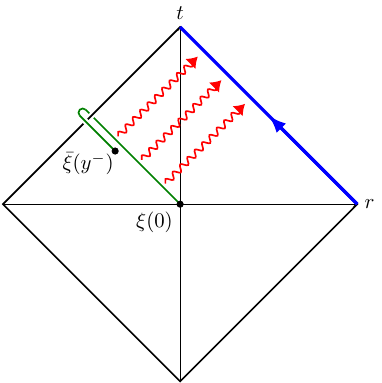}
    \caption{Penrose diagram representation of SIEC in the target fragmentation region for DIS, where all lightrays travel at 45$^\circ$. The green line represents the gauge link connecting the quark and anti-quark. The curly red lines represent collinear radiation. The blue line is the energy flow operator(s) at infinity.}
    \label{fig:penrose}
\end{figure}

\textbf{\emph{{\color{magenta}Semi-inclusive Energy Correlators.}}} We start with the operator definition of the semi-inclusive energy correlator in the target fragmentation region. The generic $n$-point energy correlator for quark is given by 
\bea\label{eq:feec}
 f_{q,n} (x,\{\Omega_i\}) &=& \int \frac{dy^- }{2\pi}  
e^{-i x y^- P^+}  
\\ 
& & \hspace{-8.ex} \times   
\langle P,\bm{s}_t|{\bar \psi}(y^-,\bm{0})  
{\cal E}(\Omega_1)  \dots {\cal E}(\Omega_n) 
\frac{\Gamma}{2}
  \psi(0) |P,\bm{s}_t \rangle \,,   \nn 
\eea
where a pictorial illustration of $f_{q,n}(x,\{\Omega_i\})$ for the DIS process is shown in Fig.~\ref{fig:penrose}. 
The key ingredient in the definition is the energy flow operator~\cite{Sveshnikov:1995vi,Tkachov:1995kk, Korchemsky:1999kt,Bauer:2008dt}, which acts as
 ${\cal E}(\Omega)|X\rangle = E(\Omega)|X\rangle$ that measures the total energy $E(\Omega) \equiv \sum_hE_h\delta(\theta_h-\theta)\delta(\phi_h-\phi)$ carried by particles at specific solid angles $\Omega = (\theta,\phi)$ relative to the beam. The energy flow operator is an integrated operator along the null infinity as denoted by the blue line in Fig.~\ref{fig:penrose}. 
Here $\bm{s}_t$ denotes the possible initial proton spin and $\Gamma$ stands for the Dirac structure matrix. At the leading power $\Gamma = \{\gamma^+, \gamma^+\gamma^5, i\sigma^{+\alpha }\gamma^5 \}$. The term $\psi(x)$ is expressed as $\psi(x) = {\cal L}_{\infty}(x)\xi(x)$, where $\xi$ represents the collinear quark field, and ${\cal L}_{\infty}$ denotes the gauge link connecting the position of the quark to light-cone infinity. The direction of the gauge link is determined by the physical process~\cite{Collins:2002kn}. 
We note that the gauge links do not cancel beyond $(0,y^-)$ due to the insertion of the energy flow operators ${\cal E}(\Omega_i)$, which induces possible spin asymmetries such as the Sivers effect~\cite{supp}.
When $n = 1$, $f_{q,n}$ reduces to the nucleon EEC~\cite{Liu:2022wop}.
The evolution of $f_{q,n}$ follows the modified DGLAP evolution~\cite{Cao:2023oef} which coincides with the DGLAP evolution at leading logarithmic accuracy. The gluonic $f_{g,n}$ can be defined similarly. 

The nucleon energy correlator $f_{q,n}(x,\{\Omega_i\})$ can be measured through the weighted cross-section 
\bea\label{eq:neec} 
\Sigma_n 
= 
\frac{1}{\sigma}
\int  d\sigma \frac{E(\Omega_1)}{Q} \dots 
\frac{E(\Omega_n)}{Q} \, w_n(\Omega_i)\,, 
\eea 
where $Q$ is chosen to be the incoming proton energy $E_P$. 
The energy flows $E(\Omega_i)$ are recorded by the calorimeter in the target fragmentation region. Here we incorporate an angular distribution function $w_n(\Omega_i)$ that assigns different weights to different angles through which the energy is distributed, thus allowing a more refined analysis of the energy flow. 


When the detector is placed in the forward region with $\theta_i \ll 1$~\footnote{Small values of $\theta_i$ can be enforced by designing $w_n$ strongly diminish at large angles. In this work, when connecting the SIECs to the TMDs, we consistently assume that $w_n$ has support primarily over the small $\theta$ regime, which, for instance, can be determined by selecting events from the forward region or a jet algorithm.}, the cross-section can be factorized as~\cite{Liu:2022wop,Cao:2023oef} 
\bea\label{eq:neec-s} 
\Sigma_n 
= \int \frac{dz}{z Q^n}\hat{\sigma}_i\left(
z P_P^+ ,\mu
\right) \int   
w_n(\Omega_i)
f_{i,n} (z,\Omega_i,\mu) \,, 
\eea 
and probes the energy correlator 
$f_{i,n}(x,\{\Omega_i\})$. We note that the angular weighting is solely acting on $f_{i,n}(x,\Omega_i)$.

The concept of the nucleon energy correlator 
can be extended to the jet or current fragmentation regions, 
which introduces the {\it fragmenting energy correlator}, as depicted in Fig.~\ref{fg:sub2}. In this context, one records the energy flow at solid angles $\Omega_i$ around an identified hadron $h$ with momentum $P_h$ and  
%
%
 measures the energy-weighted cross-section
\bea\label{eq:ffeec} 
\Sigma_n 
= \frac{1}{\sigma_h}\int  d\sigma_h
\frac{E(\Omega_i)}{Q} \dots \frac{E(\Omega_n)}{Q} \,
w_n(\Omega_i)
\,,
\eea 
where $\sigma_h$ is the semi-inclusive cross section that produces the hadron $h(P_h)$. In this case, $Q$ represents the energy that initiates $h$, for instance, $Q = \sqrt{s}/2$ at lepton colliders, or the energy of a jet if $h$ is identified within the jet. 

In general, $\Omega_i$ can be an arbitrary angle with respect to $h$. What is particularly interesting is when $\theta_{i} \ll 1$, the energy-weighted cross-section takes the factorized form 
\bea\label{eq:ffeec-s} 
\Sigma_n  = \frac{1}{\sigma_h}\int \frac{dz}{z^2 Q^n} \, 
\frac{E_kd^3\sigma_i(\mu)}{d^3k} 
\int w_n(\Omega_i)
d_{i,n}(z,\{\Omega_i\},\mu)\,, \quad
\eea 
where $\sigma_i$ is the inclusive cross section to produce a parton $i$ with momentum $k=P_h/z$ fragmenting into $h$. We have suppressed possible dependence on the spin polarization. 

The factorization theorem brings us the fragmenting energy correlator, $d_{i,n}$. The $n$-point quark 
 fragmenting energy correlator is given by 
\bea\label{eq:deec} 
&&d_{q,n}(z,\{\Omega_i\}) 
=  
\frac{z}{4N_C} 
\int \frac{dy^- }{2\pi} 
e^{\frac{i}{z}  y^- P_h }  
 \\
&&\times
 \mathrm{Tr} \Big[\langle 0| \psi(y^-,\bm{0})
{\cal E}(\Omega_1) \dots 
{\cal E}(\Omega_n) 
\sum_X |hX\rangle \langle h X|
{\bar \psi}(0) |0\rangle \Gamma \Big]  \nn\,,  
\eea 
%
where 
the angles $\Omega_i$ are measured from the $h$. 
The gluonic fragmenting energy correlator can be defined similarly.

We collectively refer to the nucleon energy correlator and the fragmenting energy correlator as the {\it semi-inclusive energy correlator}. This entails the measurement of the weighted cross-section $\Sigma_n $ in Eq.~(\ref{eq:neec}) or Eq.~(\ref{eq:ffeec}). 
%
%
Notably, when the range of the measured angle is sufficiently large to fully cover the non-perturbative regime $\theta  \sim {\cal O}\left(\frac{\Lambda_{\rm QCD}}{Q} \right)$, it guarantees that the measured $\Sigma_n$ encompasses comprehensive information on the energies and propagating angles, and therefore the full kinematics of all the particles originated from non-perturbative radiation. 
As a result, kinematic distributions of those radiations, such as the TMD PDFs, TMD FFs, the beam thrust~\cite{Stewart:2010pd}, etc., can be constructed out of the SIECs $f_{i,n}(x,\{\Omega_i\})$ and $d_{i,n}(z,\{\Omega_i\})$ through $\Sigma_n$ with suitable selections of the $w_n$, up to perturbatively calculable corrections. In this sense, the SIECs serve as generating functions for various hadron kinematic distributions. As an example, we will now demonstrate how the TMMs of the TMD PDFs and TMD FFs can be extracted separately using the nucleon energy correlator and the fragmenting energy correlator, respectively.

\textbf{\emph{{\color{magenta}TMDs out of the Energy Correlators.}}} An important insight that underpins the relationship between the TMDs and the SIECs is the fact that in the collinear limit, the transverse momentum, which is the vector sum of all particles' transverse momentum $\bm{\lambda}_{i,t}$, $\bm{k}_t = \sum_i {\bm{\lambda}}_{i,t}$, can be alternatively expressed as the transverse component of the energy flow at a fixed angle, represented by $\sum_i E_i(\theta,\phi) \sin \theta \bm{n}_t = E(\Omega) \sin\theta(\cos\phi,\sin\phi)$, and is then integrated across all the directions. 

Here, let us illustrate the derivation using the generic quark TMD PDF, which is
\bea\label{eq:deftmd} 
&& q(x,\bm{k}_t) 
= \int e^{i\bm{b}_t \cdot \bm{k}_t} q(x,\bm{b}_t)
\nn \\ 
&=& \int \frac{dy^-}{2\pi} \frac{d^2\bm{b}_t }{4\pi^2}
e^{-i x y^- P^+ + i \bm{b}_t  \cdot \bm{k}_t }  
\langle P|{\bar \psi}(y^-,\bm{b}_t) \frac{\Gamma}{2}   \psi(0) |P\rangle \,.
\eea 
Throughout this work, we stick to the matrix definition of the TMDs and notation for the related distribution functions used in~\cite{Boussarie:2023izj}.  

To proceed, we insert a complete set $1 = \sum_X|X\rangle \langle X|$ into the definition and perform the translation operation in the transverse $\bm{b}_t$ direction to find 
\bea 
 q(x,\bm{k}_t)
&=& \int \frac{dy^- }{2\pi}  
e^{-i x y^- P^+}  
\langle P|{\bar \psi}(y^-,\bm{0})\nonumber \\ 
&\times & 
\sum_X  
\delta^{(2)}\left(\bm{k}_t+\sum_{i\in X} \bm{\lambda}_{i,t} \right)
|X\rangle \langle X|  \frac{\Gamma}{2}
\psi(0) |P\rangle \,,    
\eea 
where the $\delta$-function enforces the transverse momentum conservation. We can further manipulate the equation to 
\bea 
&& q(x,\bm{k}_t)
= \int \frac{dy^- }{2\pi}  
e^{-i x y^- P^+}  
\langle P|{\bar \psi}(y^-,\bm{0})  \nonumber \\ 
&\times& \sum_X 
\delta^{(2)}\left(\bm{k}_t+
\int d\theta d\phi \,
\sin\theta \, \bm{n}_{t} 
   E(\Omega)
 \right) |X\rangle \nn \\ 
&& 
 \langle X| \frac{\Gamma}{2}
  \psi(0) |P\rangle \,,  \qquad   
\eea 
where $\bm{n}_t = (\cos\phi,\sin\phi)$ is a unit vector in the azimuthal plane and we have used in the collinear approximation, $\sum_i\bm{\lambda}_i = 
\sum_i E_i\bm{n}_{t,i} \sin\theta_i = \int d\theta d\phi 
\sum_i E_i \bm{n}_{t} \sin\theta \delta(\theta-\theta_i)\delta(\phi-\phi_i)
$ and the definition of $E(\Omega)$ is given below Eq.~(\ref{eq:neec}). Making use of the property of the asymptotic energy operator that ${\cal E}(\Omega) |X\rangle = E(\Omega)|X\rangle $ allows us to carry out the summation $\sum_X |X\rangle \langle X|=1$ and find
\bea\label{eq:qwithE} 
 q(x,\bm{k}_t)
&=& \int \frac{dy^-  }{2\pi} 
e^{-i x y^- P^+}  
\langle P|{\bar \psi}(y^-,\bm{0})  \nonumber \\ 
&\times& 
\delta^{(2)}\left(\bm{k}_t+
\int d\Omega \, \bm{n}_t  \, 
{\cal E}(\Omega)  \right) 
\frac{\Gamma}{2}
 \psi(0) |P\rangle \,,   
\eea 
where we have written $d\Omega = \sin\theta d \theta d\phi$. 

Now suppose we aim to extract the moment of the TMD PDF, denoted as $M_{\alpha}^{(n)} = \int^\mu d^2 \bm{k}_t (-\bm{k}_{t,\alpha})^n  
q(x,\bm{k}_t)$, where $\bm{k}_{t,\alpha}$ is the $\alpha$ component of the transverse momentum $\bm{k}_t$, from Eq.~(\ref{eq:qwithE}), we find that
\bea
&&  M_{\alpha}^{(n)} 
=  
\int d\Omega_1 \dots d\Omega_n \, {\bm{n}}_{1,t,\alpha}
\dots 
{\bm{n}}_{n,t,\alpha}  \,  
  \\ 
&\times & 
\int \frac{dy^-}{2\pi}   
e^{-i x y^- P^+}  
\langle P|{\bar \psi}(y^-,\bm{0})  
{\cal E}(\Omega_1)  \dots {\cal E}(\Omega_n) 
\frac{\Gamma}{2}
  \psi(0) |P\rangle \,,  \nn 
\eea 
where we quickly notice that the second line is essentially the $n$-point nucleon energy correlator in Eq.~(\ref{eq:feec}). Introducing the distribution function 
\bea\label{eq:wn}  
w_n(\Omega_i) = \prod_i^n \, d\Omega_i  \, {\bm{n}}_{1,t,\alpha}
\dots 
{\bm{n}}_{n,t,\alpha} \,, 
\eea   
allows us to establish the connection between $M_{\alpha}^{(n)} $ and the $f_{i,n}$ as~\footnote{We note that all the relations, here and below, will subject to perturbatively calculable corrections. In practice, A proper definition of the TMMs of the TMDs requires a UV regulator~\cite{Qiu:2020oqr,Ebert:2022cku,Gonzalez-Hernandez:2023iso,delRio:2024vvq}.  
On the other hand, for the SIEC, one could use a small portion of particles that surrounds the nucleon/hadron to perform the angular integration $\int w_n$, which can be achieved by a jet algorithm or by choosing particles in the forward hemisphere (target region). This leads to a mismatch between the left-hand and right-hand sides of the equation but the difference can be calculated perturbatively.
} 
\bea\label{eq:fandneec} 
 M_{\alpha}^{(n)}  = \int w_n(\Omega_i)
   f_{q,n}(x,\{\Omega_i\}) 
   \,. 
\eea 
Eq.~(\ref{eq:fandneec}) indicates that, given the $w_n(\Omega_i)$, it becomes feasible to directly examine the moment $M_{\alpha}^{(n)}$ of the TMD PDFs through $f_{i,n}(x,\{\Omega_i\})$. Other moments of the TMD PDFs can be formulated similarly with appropriate $w_n$'s. 

Similarly, we can build up the relationship between the moment of the TMD FF $d(z, z \bm{k}_t )$ and the fragmenting energy correlator. For instance, we have 
\bea\label{eq:dandffec} 
\int w_n(\Omega_i) \, 
d_{i,n}(z,\{\Omega_i\}) 
=  z^2 \int d^2 \bm{k}_t 
(\bm{k}_{t,\alpha})^n
d(z,z \bm{k}_{t}) 
 \,, 
\eea 
with the angular distribution function $ w_n$ the same as in Eq.~(\ref{eq:wn}), but the angles are measured from the hadron $h$ instead of the beam. 
Here $\bm{k}_t$ is the total transverse momentum of the particles with respective to the measured hadron $h$. 

The relationship illustrated in Eq.~(\ref{eq:fandneec}) and Eq.~(\ref{eq:dandffec}) demonstrates that the TMD PDFs and the TMD FFs can be obtained individually from the measurements in Eq.~(\ref{eq:neec}) and Eq.~(\ref{eq:ffeec}), through $f_{i,n}$ and $d_{i,n}$, respectively. 
This approach offers a notable advantage compared to the conventional methods, which typically involve the convolution of two TMDs. By using SIECs to extract TMDs, it becomes possible to isolate a specific TMD distribution, as evident from Eq.~(\ref{eq:neec-s}) and Eq.~(\ref{eq:ffeec-s}) where only one correlation function arises for each $\Sigma_n$. In addition, the transverse momentum weighted distribution enhances the high momentum mechanism, typically considered unsuitable for TMD extraction. However, the relationship derived in Eq.~(\ref{eq:fandneec}) and Eq.~(\ref{eq:dandffec}) transforms it into the $n$-point energy correlation, 
preserving the sensitivity to the non-perturbative kinematics through cutting in angles,  rendering the non-perturbative TMD distribution more viable.  

So far, we have been considering $k_t^n$-weighting for integer powers $n \geq 1$. A natural extension would be to explore non-integer values of $n$, or even to consider the limit as $n$ approaches zero, potentially following the approach of the projective energy correlator presented in \cite{Chen:2020vvp}, which we reserve for future work.



\textbf{\emph{{\color{magenta}Experimental Opportunities.}}} 
 Here we exemplify this advantage by considering several potential measurements at colliders. 

{\it 1.~Unpolarized TMDs.} One of the most straightforward extractions is the determination of the transverse moment of the unpolarized TMDs. 

For instance, the TMMs of the unpolarized TMD PDF $M_{f_1}^{(n)}(x) 
= 
\int d^2 \bm{k}_t k^{n}_t f_1(x,k_t)$ can be related to the nucleon energy correlator as $M_{f_1}^{(n)}(x) = \int w(\Omega_i) \, f_{q,n}(x,\{\Omega_i\})$ with $w_n=\prod_i^n d\Omega_i \times 1$. 
The nucleon energy correlator can be 
extracted from the DIS process with forward detectors~\cite{Liu:2022wop,Cao:2023oef}.
Notably, this extraction probes directly only the targeting TMD PDF, with no other non-perturbative objects involved, as evident from the factorization theorem in Eq.~(\ref{eq:neec-s}). The partonic cross section is identical to the inclusive DIS partonic cross section~\cite{Liu:2022wop,Cao:2023oef}. 

In practice, the measurement can be carried out using fixed-target experiments such as {\tt CLAS}~\cite{CLAS:2003umf} at JLab and boosting into the Breit frame to conduct the analysis. 
The measurement may also be feasible at the future {\tt EIC}~\cite{Accardi:2012qut} and {\tt EicC}~\cite{Anderle:2021wcy}, and hadron colliders equipped with far-forward calorimeters, such as {\tt ALICE}~\cite{ALICE:2020mso}.   

Similarly, we have $z^2\int d^2 \bm{k}_{t} k_{t}^{n} D_1(z,z k_{t}) = \int w_n(\Omega_i) d_{q,n}(z,\{\Omega_i\})$ with $w_n=\prod_i^n d\Omega_i \times 1$ for the TMMs of the unpolarized TMD FF $D_1$. The fragmenting energy correlator can be derived from measuring Eq.~(\ref{eq:ffeec}) in single hadron inclusive productions, such as $e^+e^- \to h+X$, such as {\tt BESIII}~\cite{BESIII:2020nme}. 
A jet algorithm can be applied to pre-select the particles that contribute to the angular integration $\int w_n(\Omega_i) $.  This measurement promptly probes the TMD FF since other TMDs are absent. 
Additionally, this extraction can also be performed in $ep\to hX$ and $pp\to hX$, where no TMD PDFs but only collinear PDFs are present in the cross-section $\sigma_i$ in Eq.~(\ref{eq:ffeec-s}).   


{\it 2.~Sivers Effect.} 
Similar to the previous measurement,  but equipped with a transversely polarized beam, and selecting $w_1 \propto \bm{n}_{t,\alpha}$, we can probe the first moment of the spin-dependent component of the generic TMD PDFs $\int d^2\bm{k}_t \bm{k}_{t,\alpha} q(x,\bm{k}_t)$, which is usually interpreted as the average displacement of the transverse momentum of a parton inside a hadron. 

\begin{figure}[htbp]
  \begin{center}
   \includegraphics[scale=0.3]{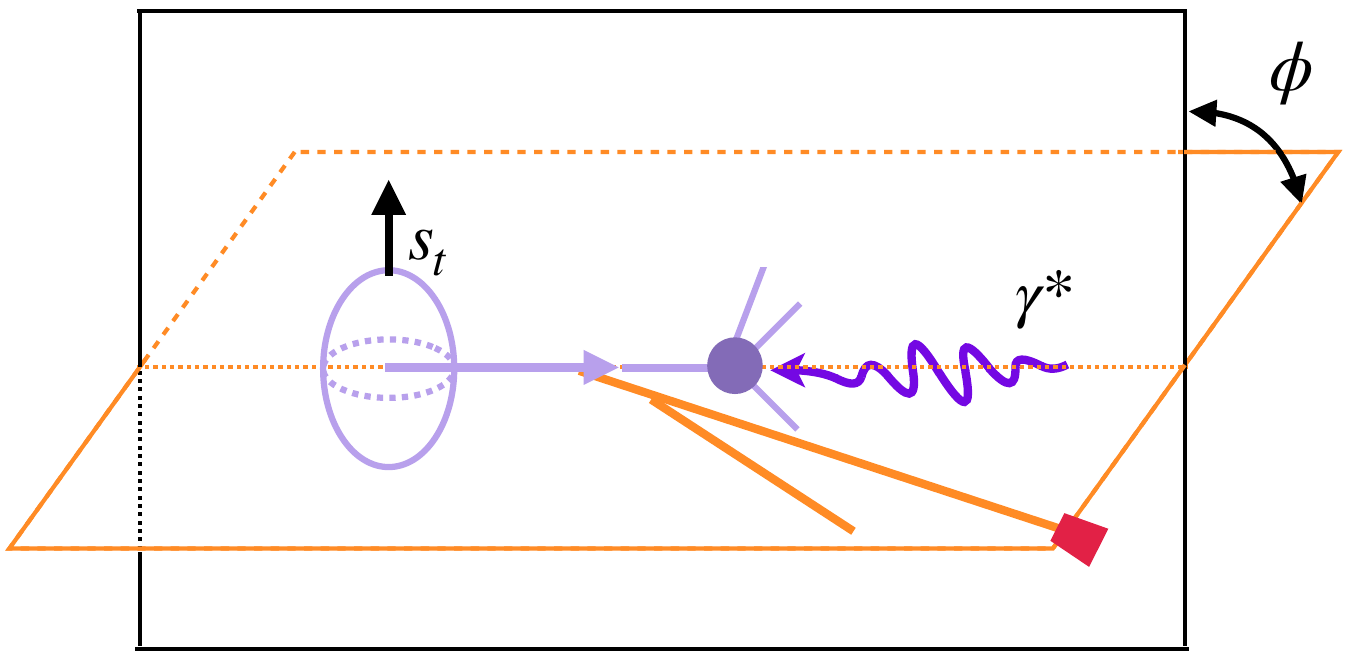} 
\caption{Sivers effect from the nucleon energy correlator in inclusive $ep^{\uparrow}$ collision, where the azimuthal angle difference $\phi$ between the forward detector (the red blob) and the polarization vector $s_t$, is measured.  }
  \label{fg:eec-sivers}
 \end{center}
\end{figure}

For instance, if we select  
%
$w_{1}^f( \Omega ) 
=  \frac{d\Omega}{\pi}
\epsilon_t^{\mu\nu}  \bm{n}_{t,\mu} \bm{s}_{t,\nu}$  
, 
we can extract the moment of the Sivers distribution $f_{1T}^\perp(x,k_t)$ through the relation
\bea 
&& \int w_1^f(\Omega) f_{q,n=1}(x,\Omega) \nn \\ 
&=& 
\int \frac{d^2\bm{k}_t}{\pi M}  
\left| \bm{k}_t\times \bm{s}_t \cdot \bm{e}_z\right|^2 f_{1T}^\perp(x,k_t) \nn \\ 
&= &  
\int \frac{dk^2_t}{2M}  k_t^2 f_{1T}^\perp(x,k_t)  \,,  
\eea  
by measuring Eq.~(\ref{eq:neec}) with one energy weight using a polarized beam. 
Fig.~\ref{fg:eec-sivers} illustrates the measurement of $\int w_1^f \, f_{q,n=1}(x,\Omega)$ in the transversely polarized DIS process. The factorization theorem satisfies Eq.~(\ref{eq:neec-s}), again, no TMDs other than the  Sivers distribution are involved, which demonstrates that the measurement of the SIEC enables the isolation and the immediate probe of the Sivers function. 

Furthermore, the first transverse moment of the Sivers function is related to the twist-$3$ Qiu-Sterman function~\cite{Qiu:1991pp} by 
$\int \frac{dk^2_t}{2M}  k_t^2 f_{1T}^\perp(x,k_t)  = \kappa \frac{1}{\pi} T(-x,0,x)$~\cite{Ji:2006ub,Ji:2006vf,Kang:2012ns,delRio:2024vvq}, therefore measuring the nucleon energy correlator offers an alternative and direct extraction of the Qiu-Sterman function. 
%



\begin{figure}[htbp]
  \begin{center}
   \includegraphics[scale=0.25]{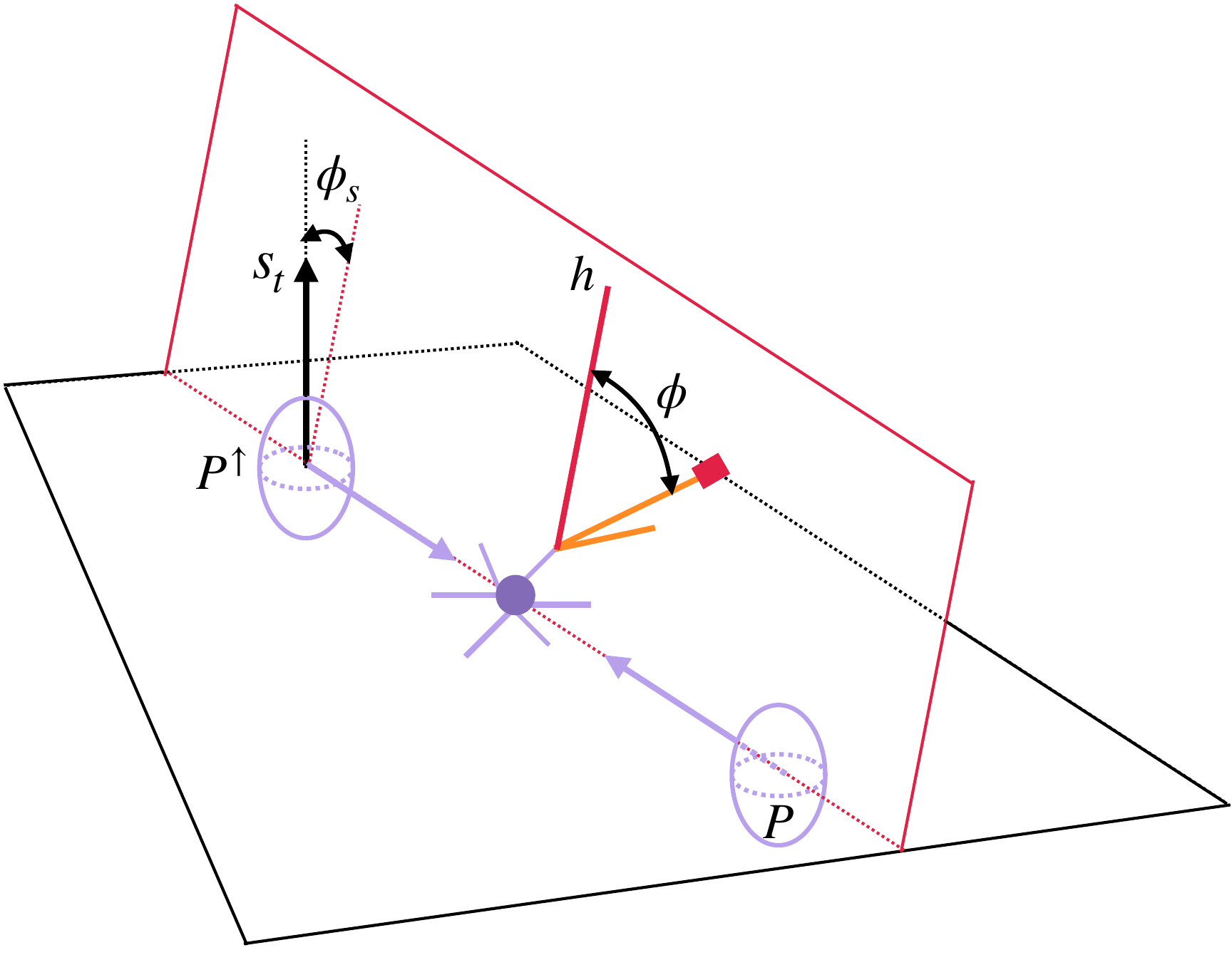} 
\caption{Collins effect from SIEC measurement in $p^{\uparrow}p$ collision, where the $\phi$'s are measured from the plane spanned by the incoming protons and the $h$.  }
  \label{fg:eec-collins}
 \end{center}
\end{figure}

{\it 3.~Collins Effect.} The transverse moment of the Collins function $H_{1T}^\perp(z,P_{h,t})$ can be extracted by measuring Eq.~(\ref{eq:ffeec}) with a transversely polarized beam. Fig.~\ref{fg:eec-collins} illustrates such a measurement in single transversely polarized $pp$ collisions. Here we choose $w_1^H = 
 \frac{d\Omega}{\pi}
\epsilon_t^{\mu\nu}  \bm{n}_{t,\mu} \bm{s}_{t,\nu}$, with all the angles measured from the identified hadron $h$, to find 
$\int w_1^H\, d_{n=1}(x,\Omega) = z^2 \int \frac{dk_t^2 }{2M} k_{t}^2 H_{1T}^\perp(z,z k_t)$. The factorization follows Eq.~(\ref{eq:ffeec-s}),   
where $d\sigma = \int d\hat{\sigma}_{ij} h_{i/P^{\uparrow}}\, f_{j/P}$ with $h_{i/P^{\uparrow}}$ represents the collinear transversity distributions in a transversely polarized proton and $f_{j/P}$ denotes the unpolarized collinear PDF. 
%


\textbf{\emph{{\color{magenta}Summary.}}}
In this work, we integrate the concept of the energy correlators~\cite{Basham:1978bw,Basham:1978zq,Hofman:2008ar,Belitsky:2013ofa,Belitsky:2013xxa,Kologlu:2019mfz,Korchemsky:2019nzm,Dixon:2019uzg,Chen:2019bpb,Chicherin:2020azt,Chen:2020adz,Chen:2020vvp,Chang:2020qpj,Li:2021zcf,Jaarsma:2022kdd,Komiske:2022enw,Holguin:2022epo,Yan:2022cye,Chen:2022jhb,Chang:2022ryc,Chen:2022swd,Lee:2022ige,Larkoski:2022qlf,Ricci:2022htc,Yang:2022tgm,Andres:2022ovj,Craft:2022kdo,Chen:2022pdu,Andres:2023xwr,Devereaux:2023vjz,Andres:2023ymw,Jaarsma:2023ell,Lee:2023npz,Lee:2023tkr,Yang:2023dwc,Chen:2023zzh,Holguin:2023bjf,Cao:2023qat,Gao:2023ivm,Yang:2024gcn} into hadron structure studies by introducing the semi-inclusive energy correlators (SIECs) that encompass the nucleon energy correlator~\cite{Liu:2022wop} and the fragmenting energy correlator. The SIECs incorporate comprehensive information on the radiations surrounding the hadron, enabling a detailed analysis of the hadron's intrinsic structures. We demonstrate the effectiveness of this probe by showcasing how the extraction of the transverse momentum moments (TMMs) of TMD distributions can be achieved through the SIECs with suitable angular weights. 

One immediate advantage of this approach to the TMDs 
is its capability to pinpoint one specific TMD distribution for each measurement, providing a direct means of examining the distribution, 
while many conventional TMD studies, including recent proposals using EEC~\cite{Moult:2018jzp,Gao:2019ojf,Li:2020bub,Gao:2023ivm,Kang:2023big}, enforce the back-to-back configuration and involve two TMDs.  
Additionally, the semi-inclusive correlators convert the TMMs into the $n$-point functions, avoiding the enhancement of the high transverse momentum mechanism, thereby making the extraction of the non-perturbative TMD distribution more feasible. 
We thus anticipate the SIECs open up new avenues for the exploration of the TMD structure functions. 

Although this manuscript has focused explicitly on the extraction of TMDs through SIECs, the application of the SIECs extends beyond TMD studies. The hadronic transverse and spin structures can be probed directly by just looking at the $n$-point angular correlation of the SIECs, without resorting to other observables. 
One can also measure the energy flows with additional information on quantum numbers ~\cite{Chicherin:2020azt,Li:2021zcf,Jaarsma:2023ell,Lee:2023npz,Andres:2023ymw}, to allow us additional access to the distribution of the particles carrying the specific quantum charges around the target hadron.
Furthermore, if one integrates over $P_h^+$ and sums over hadrons $h$ with the same quantum number $ q_h$ in Eq.~(\ref{eq:deec}), one essentially counts the number of $h$ with $q_h$ propagating along the same direction $\hat{n}$. One thus can relate 
\bea \sum_{h\in q_h}
\int dP_h^+ \,  \, 
d_{q,n}(z,\{\Omega_i\}) 
\sim  
\langle {\cal E}(\Omega_1) \dots {\cal E}(\Omega_n) {\cal Q}_h(\hat{n}) \rangle  \,. 
\qquad 
\eea 
We refer the readers to the Supplemental Materials~\cite{supp} for the derivation and notations. Here
 ${\cal Q}_h(\hat{n}) 
$
is the light-ray operator that measures the ``charge" flux constructed out of the interpolating current $J_h^\mu(x)$ acting on the space with quantum number $q_h$. When $n=0$, it reduces to the first moment of the track function in its general mean~\cite{Li:2021zcf,Jaarsma:2022kdd}. The connection between the SIECs and the correlators with currents~\cite{Belitsky:2013xxa,Chicherin:2020azt} and the track function may provide new prospects to study these objects.
Therefore, with access to comprehensive information about radiation, we believe the SIECs are poised to offer us a fresh revelation of the hadronic three-dimensional structure. 


\begin{acknowledgments}
 \textbf{\emph{Acknowledgement.}} 
 We are grateful to Jian-Ping Ma, Dingyu Shao and Feng Yuan for their useful discussions.
X.~L. is supported by the Natural Science Foundation of China under contract No.~12175016 (X.~L.). H.~X.~Z. is supported by the start-up fund of Peking University and the Asian Young Scientist Fellowship.

\end{acknowledgments}




\bibliographystyle{h-physrev}   
\bibliography{refs}

\clearpage



\section*{Supplementary material}

# Supplemental Materials for "TMDs from Semi-inclusive Energy Correlators"

Xiaohui Liu[1,2,*] and Hua Xing Zhu[3,4,†]

[1]*Department of Physics, Beijing Normal University, Beijing, 100875, China*
[2]*Key Laboratory of Multi-scale Spin Physics, Ministry of Education,
Beijing Normal University, Beijing 100875, China*
[3]*School of Physics, Peking University, Beijing, 100871, China*
[4]*Center for High Energy Physics, Peking University, Beijing 100871, China*
(Dated: March 13, 2024)

## I. SIVERS EFFECT IN SIEC

Here we study the Sivers effect of the semi-inclusive energy correlator (SIEC). The definition of the target region SIEC, or the nucleon energy correlator, with one insertion of the energy flow operator $\mathcal{E}$ for DIS and Drell-Yan is given by

$$f^{\rm DIS}_{q,n=1}(x,\Omega) = \int \frac{dy^-}{2\pi} e^{-ixy^-P^+} \langle P\boldsymbol{s}_t|\bar{\xi}(y^-,\mathbf{0})\mathcal{L}(y^-,\infty)\mathcal{E}(\Omega)\frac{\gamma^+}{2}\mathcal{L}(\infty,0)\xi(0)|P\boldsymbol{s}_t\rangle \,, \qquad (1)$$

and

$$f^{\rm DY}_{q,n=1}(x,\Omega) = \int \frac{dy^-}{2\pi} e^{-ixy^-P^+} \langle P\boldsymbol{s}_t|\bar{\xi}(y^-,\mathbf{0})\mathcal{L}(y^-,-\infty)\mathcal{E}(\Omega)\frac{\gamma^+}{2}\mathcal{L}(-\infty,0)\xi(0)|P\boldsymbol{s}_t\rangle \,, \qquad (2)$$

respectively. We explicitly show the direction of the gauge link. We note that the gauge link outside $(0, y^-)$ does not cancel due to the insertion of $\mathcal{E}(\Omega)$ at $y^+ = \infty$. We illustrate $f_{q,n=1}(x,\Omega)$ for both the DIS and the Drell-Yan in Fig. 1. Any gluons emitted from the gauge link $\mathcal{L}$ at finite $y^-$ reaching the detector $\mathcal{E}(\Omega)$ would have momentum $p^- \ll p^+$, typical of collinear gluons. However, for emissions near the boundary of the Penrose diagram, from $y^- \to \infty$ to $y^+ = \infty$, their momentum will tend to $p^+, p^- \to 0$. This could generate the Glauber contribution to the energy correlator, leading to spin asymmetry.

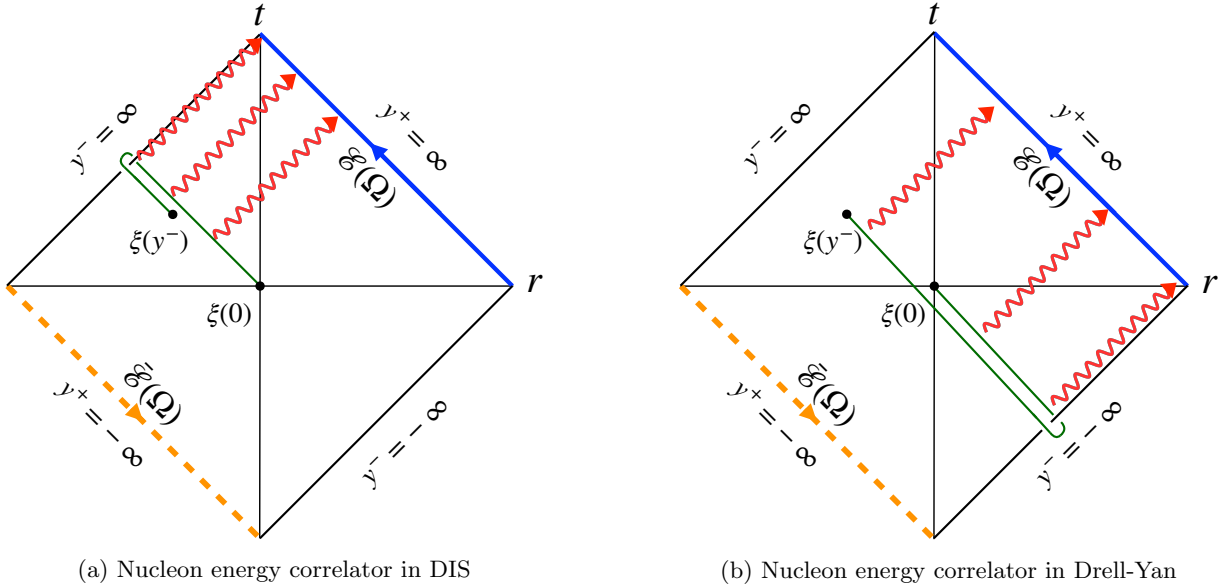

(a) Nucleon energy correlator in DIS    (b) Nucleon energy correlator in Drell-Yan

FIG. 1: Penrose diagrams for the nucleon energy correlator in DIS and Drell-Yan.

---

[*] xiliu@bnu.edu.cn
[†] zhuhx@pku.edu.cn

To proceed, we decompose the transverse spin state $|P\boldsymbol{s}_t\rangle$ by the helicity states of the incoming proton $|P\boldsymbol{s}_t\rangle = \sum_{\alpha=L,R} c_\alpha |P\alpha\rangle$ such that

$$f_{q,n=1}^{\text{DIS/DY}}(x,\Omega) = \sum_{\alpha,\alpha'=L,R} \rho_{\alpha'\alpha} f_{q,n=1,\alpha\alpha'}^{\text{DIS/DY}}(x,\Omega), \tag{3}$$

where $\rho_{\alpha'\alpha} = c_{\alpha'} c_\alpha^*$ and $f_{q,n=1,\alpha\alpha'}^{\text{DIS/DY}}(x,\Omega)$ is the nucleon energy correlator in the helicity basis obtained by replacing $|P\boldsymbol{s}_t\rangle$ with $|P\alpha\rangle$ in Eq. (1) and Eq. (2).

Now following Collins [1], it is straightforward to show by the symmetry argument under the joint parity and time reversal transformation $\mathcal{PT}$ that

$$f_{q,n=1,LR}^{\text{DIS}}(x,\Omega) = -f_{q,n=1,LR}^{\text{DY}}(x,\Omega), \qquad f_{q,n=1,RL}^{\text{DIS}}(x,\Omega) = -f_{q,n=1,RL}^{\text{DY}}(x,\Omega), \tag{4}$$

where the gauge link switches directions under $\mathcal{PT}$. To see this, we note that majority parts within $f_{q,n=1,\alpha\alpha'}$ are similar to the TMDs considered in [1], with the exception of the energy flow operator

$$\mathcal{E}(\Omega) = \int_0^\infty dt \lim_{r\to\infty} r^2 n_i T^{0i}(t, rn_i). \tag{5}$$

Here $T^{0i}$ is the energy flux component of the energy stress tensor $T^{\mu\nu}$ and transforms under $\mathcal{PT}$ as

$$(\mathcal{PT})^\dagger T^{0i}(x)(\mathcal{PT}) = T^{0i}(-x). \tag{6}$$

The transform form can be readily derived. For instance, the quark stress tensor contains

$$T^{0i}(x) \propto i\bar{\psi}\gamma^0 \mathcal{D}^i \psi(x) + i\bar{\psi}\gamma^i \mathcal{D}^0 \psi(x) + \ldots, \tag{7}$$

while we know that

$$\mathcal{P}^\dagger \bar{\psi}\gamma^\mu \psi(x) \mathcal{P} = (-1)^\mu \bar{\psi}\gamma^\mu \psi(t, -\boldsymbol{x}), \qquad \mathcal{T}^\dagger \bar{\psi}\gamma^\mu \psi(x) \mathcal{T} = (-1)^\mu \bar{\psi}\gamma^\mu \psi(-t, \boldsymbol{x}), \tag{8}$$

where $(-1)^\mu = 1$ for $\mu = 0$ and $-1$ for $\mu = 1, 2, 3$, therefore

$$(\mathcal{PT})^\dagger \bar{\psi}\gamma^\mu \psi(x)(\mathcal{PT}) = \bar{\psi}\gamma^\mu \psi(-x) \tag{9}$$

By further noting that $\mathcal{T}i = -i\mathcal{T}$, as well as $(\mathcal{PT})^\dagger \partial^\mu (\mathcal{PT}) = -\partial^\mu$, we arrive at Eq. (6).

Therefore, under the $\mathcal{PT}$ operation, the energy flow operator transforms as

$$(\mathcal{PT})^\dagger \mathcal{E}(\mathcal{PT}) = \int_0^\infty dt \lim_{r\to\infty} r^2 n_i T^{0i}(-t, -rn_i) = \int_0^{-\infty} dt \lim_{r\to\infty} r^2 (-n_i) T^{0i}(t, -rn_i) \equiv \bar{\mathcal{E}}, \tag{10}$$

where $\bar{\mathcal{E}}$ records the energy flow along $-n_i$ direction at $r \to \infty$ from $t = 0$ to the infinite past.

In addition, the $f_{q,n=1}$ should not change with the replacement

$$\mathcal{E} \to \mathcal{E} + \bar{\mathcal{E}}, \tag{11}$$

since the insertion of $\bar{\mathcal{E}}$ at infinite past contributes vanishingly to the matrix in Eq. (1) and Eq. (2), which can be seen from the Penrose diagram in Fig. 1. Apparently, $\mathcal{E} + \bar{\mathcal{E}}$ is invariant under $\mathcal{PT}$ and the rest part of the matrix element behaves the same under $\mathcal{PT}$ as the TMD Sivers effect [1]. Therefore one immediately concludes Eq. (4).

One can decompose Eq. (1) and Eq. (2) by spin structures. The most generic decomposition [2] is

$$f_{q,n=1}(x,\Omega) = f_{q,n=1}^{un.}(x,\Omega) + \kappa \epsilon_t^{\mu\nu} s_{t,\mu} n_{t,\nu} \Delta f_{q,n=1}(x,\Omega) \tag{12}$$

where $\kappa_{\text{DY}} = -\kappa_{\text{DIS}}$. Here $f_{q,n=1}^{un.}(x,\Omega)$ is the unpolarized energy correlator and $\Delta f_{q,n=1}(x,\Omega)$ encodes the Sivers effect.

## II. DERIVATION OF EQ. (16)

We sum over all hadrons belonging to the same quantum number $q_h$ along a direction $n$. The quark fragmenting energy correlator then becomes

$$D_{q,n}(z, \{\Omega_i\}) = \frac{z}{2N_C} \sum_{r=1, h\in q_h}^{r=\infty} \int \frac{dy^-}{2\pi} e^{iy^- P/z} \text{Tr}\left[\langle 0|\psi(y^-)\mathcal{E}_1 \cdots \mathcal{E}_n |\underbrace{h \cdots h}_{r} X\rangle\langle X \underbrace{h \cdots h}_{r}|\bar{\psi}|0\rangle \frac{\Gamma}{2}\right]. \tag{13}$$

We note that $\int dz D_{q,n}(z,\{\Omega_i\})$ is the same as $\sum_{h\in q_h} \int dz\, d_{q,n}(z,\{\Omega_i\})$ that counts the number of hadrons $h$ with quantum number $q_h$ fragmentated from a parent quark. Here we try to argue the relation among $\int dz\, D_{q,n}(z,\{\Omega_i\})$, the correlation function with currents and track functions [3, 4].

For simplicity, we consider mesons $h$ made up of quarks with one flavor. The interpolating annihilation operator for the meson can be chosen as [5]

$$a_{h,P_h} = \sqrt{\frac{P_h}{2(2\pi)^3}} \int dx\, \Psi_h(x) \frac{1}{\sqrt{N_C}} \sum_c b^c_{xP_h} d^c_{-(1-x)P_h} \tag{14}$$

where $b$ and $d$ are annihilation operators for the valence quark and anti-quark moving along the direction of the meson $h$. Here $c$ denotes the color index and we have made the summation over the color explicity. The wave function $\Psi_h(x)$ is complete on the sub-space with quantum number $q_h$ such that

$$\sum_{h\in q_h} \Psi^\dagger(y)_h \Psi_h(x) = \delta(x-y) \mathcal{P}_{q_h}, \tag{15}$$

where we sum over all possible states with the same quantum number $q_h$. $\mathcal{P}_{q_h}$ is the projection operator that picks the sub-space with the quantum number $q_h$.

Now we consider

$$\int dP_h P_h \sum_{h\in q_h} a^\dagger_{h,P_h} a_{h,P_h} = \frac{1}{N_C} \int dP_h\, P_h \int dx \int dy \sum_{h\in q_h} \frac{P_h}{2(2\pi)^3} \Psi^\dagger_h(y)\Psi_h(x) \sum_{c,c'} d^{c'\dagger}_{-(1-y)P_h} b^{c'\dagger}_{yP_h} b^c_{xP_h} d^c_{-(1-x)P_h}$$

$$= \frac{1}{N_C} \int \frac{dP_h}{2(2\pi)^3}\, P_h \int dx P_h \sum_{cc'} d^{c'\dagger}_{-(1-x)P_h} b^{c'\dagger}_{xP_h} b^c_{xP_h} d^c_{-(1-x)P_h} \mathcal{P}_{q_h}$$

$$= \frac{1}{N_C} \sum_{cc'} \left( \int dp\, p\, b^{c'\dagger}_p b^c_p \int \frac{dq}{2(2\pi)^3} d^{c'\dagger}_{-q} d^c_{-q} + \int \frac{dp}{2(2\pi)^3} b^{c'\dagger}_p b^c_p \int dq\, q\, d^{c'\dagger}_{-q} d^c_{-q} \right) \mathcal{P}_{q_h}$$

$$\to \frac{1}{2}\sum_c \left( \int dp\, p\, b^{c\dagger}_p b^c_p + \int dq\, q\, d^{c\dagger}_{-q} d^c_{-q}\right) \mathcal{P}_{q_h} = \frac{1}{2} \sum_c \int dp\, p\, (b^{c\dagger}_p b^c_p - d^{c\dagger}_p d^c_p) \mathcal{P}_{q_h}, \tag{16}$$

in the second line, we have used the completeness condition for $\Psi_h$; in the third line, we have let $p = xP_h$ and $q = (1-x)P_h$. The replacement in the last line is valid because the operator will act on a physical (color-singlet) state $|X\rangle$ and only colorless $q\bar{q}$ pairs can occur in $h$ such that $\mathcal{P}_{q_h}|X\rangle \sim \sum_{h\in h_q} |h(q^c(k)\bar{q}^c(k')); X'\rangle$ [5]. The result is not surprising, it simply indicates that, on the $q_h$ subspace, counting the number of hadrons with momentum $P_h$ is equivalent to counting the number of valance quarks along the direction of the hadron.

Now by using [6]

$$\frac{1}{4}\sum_c \int \frac{dp}{2(2\pi)^3} p\, (b^{c\dagger}_p b^c_p - d^{c\dagger}_p d^c_p) \mathcal{P}_{q_h} = \int_{-\infty}^{\infty} dx^- \lim_{x^+\to\infty} \left(\frac{x^+}{2}\right)^2 \frac{\bar{n}_\mu}{2} J^\mu(x) \mathcal{P}_{q_h} \equiv \mathcal{Q}_h(\hat{n}), \tag{17}$$

where $J^\mu$ is the EM current $J^\mu(x) = \bar\psi \gamma^\mu \psi(x)$, $\hat{n}$ is along the direction of hadrons $h$, $v^- \equiv \hat{n}\cdot v$ and $v^+ = \bar{\hat{n}}\cdot v$, we arrive at Eq. (16)

$$\frac{1}{P}\int dP_h\, D_{q,n}(z,\{\Omega_i\}) = \frac{32\pi^3}{P^2} \langle \mathcal{E}(\Omega_1)\ldots\mathcal{E}(\Omega_n) \mathcal{Q}_h(\hat{n})\rangle, \tag{18}$$

where $z = \frac{P_h}{P}$, $P$ is the parent quark momentum, and

$$\langle \mathcal{E}(\Omega_1)\ldots \mathcal{E}(\Omega_n) \mathcal{Q}_h(\hat{n})\rangle = \frac{1}{2N_c} \int \frac{dy^-}{2\pi} e^{iy^- P} \langle 0|\psi(y^-,\mathbf{0})\mathcal{E}(\Omega_1)\ldots\mathcal{E}(\Omega_n)\mathcal{Q}_h(\hat{n})\Gamma\bar\psi(0)|0\rangle, \tag{19}$$

The derivation extends to multi-flavors and higher moments. For instance, for $\pi$'s, one can choose $J^\mu = (\bar{u}\gamma^\mu u - \bar{d}\gamma^\mu d)/2$ in the iso-spin limit. One can also show that

$$\frac{1}{P^2}\int dP_h P_h D_{q,n}(z,\{\Omega_i\}) = \frac{64\pi^3}{P^3} \langle \mathcal{E}(\Omega_1)\ldots\mathcal{E}(\Omega_n)(\mathcal{T}_h(\hat{n}) + \mathcal{Q}_h(\hat{n})\mathcal{Q}_h(\hat{n}))\rangle \tag{20}$$

where

$$\mathcal{T}_h(\hat{n}) = \int_{-\infty}^{\infty} dx^- \lim_{x^+\to\infty} \left(\frac{x^+}{2}\right)^2 \bar\psi(x) \frac{\gamma^+}{2} (\frac{i}{2} D^+) \psi(x) \mathcal{P}_{q_h}. \tag{21}$$

Last we note that when $n = 0$, it reduces to the track function in its general mean [4, 7], such that $\int dz \, D_{q,0}(z) = T_q(1)$, $\int dz \, z \, D_{q,0}(z) = T_q(2)$, so on and so forth, or

$$\langle \mathcal{Q}_h \rangle \sim \langle \sigma_q(1) \rangle, \qquad \langle \mathcal{T}_h \rangle \sim \langle \sigma_q(2) \rangle, \qquad \ldots \tag{22}$$

where $\sigma_q(n)$ is the central moments [7].

---



\end{document}